\documentstyle[11pt,aaspp4]{article}

\begin{document}

\title{
	A blind test of photometric redshift prediction\altaffilmark{1}
}
\author{
	David~W.~Hogg\altaffilmark{2,3,4},
	Judith~G.~Cohen\altaffilmark{5},
	Roger~Blandford\altaffilmark{2},
	Stephen~D.~J.~Gwyn\altaffilmark{6},
	F.~D.~A.~Hartwick\altaffilmark{6},
	B.~Mobasher\altaffilmark{7},
	Paula~Mazzei\altaffilmark{8},
	Marcin~Sawicki\altaffilmark{9},
	Huan~Lin\altaffilmark{9},
	H.~K.~C.~Yee\altaffilmark{9},
	Andrew~J.~Connolly\altaffilmark{10},
	Robert~J.~Brunner\altaffilmark{10},
	Istvan~Csabai\altaffilmark{10,11},
	Mark~Dickinson\altaffilmark{10,12,13},
	Mark~U.~SubbaRao\altaffilmark{10},
	Alexander~S.~Szalay\altaffilmark{10},
	Alberto Fern\'andez-Soto\altaffilmark{14,15},
	Kenneth M. Lanzetta\altaffilmark{15},
	Amos Yahil\altaffilmark{15}
}
\altaffiltext{1}{
Based on observations taken with the NASA/ESA Hubble Space Telescope,
which is operated by AURA under NASA contract NAS 5-26555; at the
W. M. Keck Observatory, which is operated jointly by the California
Institute of Technology and the University of California; and at the
Kitt Peak National Observatory, which is operated by AURA under
cooperative agreement with the NSF.
}
\altaffiltext{2}{
Theoretical Astrophysics, California Institute of Technology,
mail code 130-33, Pasadena, CA 91125, USA;
{\tt hogg, rdb@tapir.caltech.edu}
}
\altaffiltext{3}{
Present address:
School of Natural Sciences, Institute for Advanced Study,
Olden Lane, Princeton NJ 08540, USA;
{\tt hogg@ias.edu}
}
\altaffiltext{4}{
Hubble Fellow.
}
\altaffiltext{5}{
Palomar Observatory, California Institute of Technology,
mail code 105-24, Pasadena, CA 91125, USA;
{\tt jlc@astro.caltech.edu}
}
\altaffiltext{6}{
Department of Physics and Astronomy, University of Victoria,
Box 3055, Victoria, British Columbia V8W 3P6, Canada;
{\tt gwyn, hartwick@uvastro.phys.uvic.ca}
}
\altaffiltext{7}{
Astrophysics Group, Blackett Laboratory, Imperial College,
Prince Consort Road, London SW7~2BZ, UK;
{\tt b.mobasher@ic.ac.uk}
}
\altaffiltext{8}{
Osservatorio Astronomico,
Vicolo dell'Osservatorio, 5, 35122 Padova, Italy;
{\tt mazzei@astrpd.pd.astro.it}
}
\altaffiltext{9}{
Department of Astronomy, University of Toronto,
Toronto, Ontario M5S 3H8, Canada;
{\tt sawicki, lin, hyee@astro.utoronto.ca}
}
\altaffiltext{10}{
Department of Physics and Astronomy, The Johns Hopkins University,
Baltimore, MD 21218, USA;
{\tt ajc, rbrunner, csabai, subbarao, szalay@pha.jhu.edu}
}
\altaffiltext{11}{
Department of Physics, E\"{o}tv\"{o}s University,
Budapest, Hungary, H-1088
}
\altaffiltext{12}{
Alan C. Davis Fellow
}
\altaffiltext{13}{
Space Telescope Science Institute,
3700 San Martin Drive, Baltimore, MD 21218, USA;
{\tt med@stsci.edu}
}
\altaffiltext{14}{
Present address: Department of Astrophysics and Optics, 
School of Physics, University of New South Wales,
Sydney, NSW-2052, Australia;
{\tt fsoto@edwin.phys.unsw.edu.au}
}
\altaffiltext{15}{
Department of Physics and Astronomy,
State University of New York at Stony Brook, Stony Brook, NY 11794-2100;
{\tt lanzetta, ayahil@sbast4.ess.sunysb.edu}
}

\newpage
\begin{abstract}
Results of a blind test of photometric redshift predictions against
spectroscopic galaxy redshifts obtained in the Hubble Deep Field with
the Keck Telescope are presented.  The best photometric redshift
schemes predict spectroscopic redshifts with a redshift accuracy of
$|\Delta z|<0.1$ for more than 68~percent of sources and with $|\Delta
z|<0.3$ for 100~percent, when single-feature spectroscopic redshifts
are removed from consideration.  This test shows that photometric
redshift schemes work well at least when the photometric data are of
high quality and when the sources are at moderate redshifts.
\end{abstract}


\section{Introduction}

Photometric redshift prediction techniques, in which galaxy redshifts
are estimated using only broad-band photometric information, promise
the acquisition of large numbers of redshifts of very faint sources
with comparatively little telescope time, at least relative to
performing a spectroscopic survey to comparable depth.  The idea dates
back to Baum (1962) who used nine-band photoelectric photometry to
(essentially) locate the 4000\,\AA\ break in elliptical galaxies.  Loh
\& Spillar (1986) generalized this technique, with 6-band CCD
photometry, to apply to a wide range of galaxy types.  With only four
photographic bands, Koo (1985) was able to estimate redshifts from
colors after plotting lines of constant redshift on color-color plots.

In the astronomical community, there is some perceived skepticism that
photometric redshift estimation schemes are reliable or have been
tested fairly.  This skepticism may not appear in the literature or be
well-justified by demonstrated failings in photometric redshift
techniques; however we feel that it is strong enough to motivate a
blind test of the methods.  In this paper, the results of such a test
are presented.  The test was administered by three of us (DWH, JGC,
RB) with no stake in the outcome.

For data, the imaging and spectroscopy of the Hubble Deep Field (HDF,
Williams et al 1996) was chosen.  The reasons for this include that it
has been imaged to enormous depth in the $F300W$, $F450W$, $F606W$,
and $F814W$ bandpasses (close to $U$, $B$, $V$ or $R$, and $I$) with
the Hubble Space Telescope (HST), that it has been imaged from the
ground in the near-infrared $J$, $H$ and $K$ bandpasses (Hogg et al
1997; Dickinson 1997), and that there are various spectroscopic
surveys underway in the field (Cohen et al 1996; Steidel et al 1996;
Lowenthal et al 1997) which can be used to train or develop
photometric prediction methods.

There are three respects in which this is not the most stringent test
on the photometric redshifts.  The first is that the number of sources
is small.  Only 27 new sources in the HST image of the HDF had
spectroscopic observations by the time the blind test was announced,
so only this sample was available for the test.  After reduction and
analysis, only 21 of these have high-quality redshift determinations.
The second is that the observational concentration on the HDF has been
so intense that the photometric database is of much higher quality
than that of any other existing field or any likely to exist in the
near future.  The third is that the spectroscopic redshifts used in
the blind test are taken from a magnitude limited survey of normal
galaxies (Cohen et al 1996; 1997) underway with the LRIS instrument
(Oke et al 1995) on the Keck Telescope.  This survey, like most of its
type, has difficulty identifying sources with redshifts in the range
$1.5<z<2.5$ because of the lack of strong spectral features in the
visual region free of strong night sky emission.  So the present work
cannot test photometric redshift prediction techniques in this crucial
redshift range, where perhaps they are most promising as a
revolutionary tool in cosmology.  On the other hand, this test is in
one respect more stringent than previous tests: the sources employed
for this blind test are on average fainter than those in the
``training set,'' which is primarily the 1996 HDF redshifts (Cohen et
al 1996).

\section{Data and blind-test procedure}

During the 1997 February 6 and 7 runs on the Keck~II Telescope, 27
sources in the HST-imaged region of the HDF were observed
spectroscopically with the LRIS instrument as part of the survey of
Cohen et al (1996, 1997).  The positions but not redshifts of the 27
sources were announced by email to the HDF follow-up community on 1997
March 17, along with a call for photometric redshift predictions.  The
sources are listed in Table~\ref{tab:sources}, along with chip numbers
(in the HST HDF images), $x,y$ positions (in the ``Version 2''
reductions of the HST HDF data), magnitudes and spectroscopic
redshifts $z$ based on a fast-pass reduction of the spectra.  These
and other HDF redshifts, along with details of selection and
observing, will be published in Cohen et al (1997).  Briefly, the
sources all have $R<24$~mag and were spectroscopically observed with
the same instrument setup as that described in Cohen et al (1996).
The magnitudes in Table~\ref{tab:sources} are Vega-relative and
measured through 1.5~arcsec diameter apertures centered on the
sources; they are provided only as rough flux estimates.  Two sources
have redshifts based on a single emission line assumed to be the
[O\,II] 3727\,\AA\ line, and two have redshifts based on a weak break
assumed to be the 4000\,\AA\ break.  These four spectroscopic
redshifts are deemed less certain and indicated in
Table~\ref{tab:sources} and the results.  Details of the spectroscopic
analysis will be presented by Cohen et al (1997).

By the close of the deadline on 1997 April 25, six
photometric-redshift groups had submitted predictions.  After the
results were presented at a meeting in 1997 May (Hogg, Cohen \&
Blandford 1997), five groups agreed to participate in this summary
paper.

\section{Prediction schemes}

The photometric prediction schemes for the participating groups are
described in this Section, listed in the order in which the entries to
the blind test were submitted.

\subsection{Victoria}

The Victoria method (Gwyn \& Hartwick 1996) uses the standard template
fitting method described in Gwyn (1997), although with modified
templates.  A spectral energy distribution (SED) is derived from the
$UBRI$ photometry of the HDF for each galaxy.  This SED is compared to
a series of redshifted templates.  Two sets of templates are used: The
model spectra of Bruzual \& Charlot (1993, hereafter BC) and the
empirical spectra of Coleman, Wu \& Weedman (1980, hereafter CWW, as
extended by Ferguson \& McGaugh 1994).  The former (BC) spectra have
the advantage that they extend further into the UV and consequently
can be used at higher redshifts.  The latter (CWW) spectra were found
to produce more accurate redshifts where they can be used (that is at
low redshift).  Both sets of templates are corrected for intergalactic
absorption using the prescription of Madau (1995).  Use of the CWW
templates and correction for intergalactic absorption represent two
important improvements over the technique employed in Gwyn \& Hartwick
(1996).  As a first cut, a photometric redshift is determined by
comparing the observed SED to the redshifted BC templates.  Then, if
this redshift is $z<1.5$, it is discarded in favor of a photometric
redshift determined using the CWW templates.

\subsection{Imperial College}

The photometric redshifts of the Imperial College group are estimated
by comparing observed SEDs with model SEDs for different populations
of galaxies, shifted to different redshifts.  This is performed in the
following steps: (1)~The observed SEDs of individual galaxies are
established, using the U, B, V and I-band observations of the HDF. The
U-band data is crucial for locating the 4000\,\AA\ break.  (2)~Model
SEDs corresponding to 4 template spectra (elliptical, spiral,
irregular and starbursts) are constructed following recipes described
by Mazzei et al (1994; 1995).  These SEDs are extended from the far
ultraviolet to 1~${\rm \mu m}$ wavelengths and incorporate stellar
emission, internal extinction and re-emission by dust. At the
rest-frame these fit the observed SEDs of local galaxies of their
respective type. These models, computed using the evolutionary
population synthesis technique, assume a star formation rate and the
initial mass function, permitting calculation of the template SEDs at
different redshifts.  The model parameters are then constrained by
fitting the SEDs of a sample of 53 galaxies in the HDF with
spectroscopic redshifts (Cohen et al 1996; Steidel et al 1996).  The
rms scatter between the photometric redshifts and their spectroscopic
counterparts (Cohen et al 1996; Steidel et al 1996) is found to be
0.11.  The final models correspond to formation redshifts $z_f=5$
(13~Gyr) for ellipticals, 2 (10--11~Gyr) for spirals, 1
(0.8--0.9~Gyrs) for irregulars and 5 (13~Gyr) for starbursts (Taking
$H_0=50~{\rm km\,s^{-1}\,Mpc^{-1}}$ and $q_0=0.5$). A detailed
description of the model SEDs is given in Mobasher et al (1997).
(3)~A matrix with elements consisting of the template SEDs for the 4
types of galaxies, shifted in redshift space, is generated with the
evolutionary models. The redshift and spectral type corresponding to
the template SED closest to the observed SED are associated with each
galaxy.  A detailed description of the technique is given in Mobasher
et al (1996).

In the original blind test, because of a coordinate system
discrepancy, the list of sources submitted by the Imperial College
group did not match the list presented in the call for entries to the
test.  After the results were presented, the Imperial College group
submitted photometric redshifts for the correct list of sources.  In
principle this compromises the blindness of the test although no
modification was made to the Imperial College photometric redshift
prediction technique between the first and second submissions.

\subsection{Toronto 4-color}

The Toronto 4-color predictions are the redshift estimates of Sawicki
et al (1997).  This method uses the 4 HST filters to measure the
colors of each galaxy, and automatic morphological classification to
weed out stars.  Redshifts are determined by comparing the observed
colors with those of a set of model templates.  The templates were
computed on the basis of the empirical $z \sim 0$ spectral energy
distributions of CWW; these SEDs have been augmented and extended into
the UV by grafting on the spectral shapes of Bruzual \& Charlot
(1993); they were then UV-supressed using Madau's (1995) prescription
for high-redshift intergalactic H continuum and line blanketing; the
resultant SEDs were convolved with instrumental response curves to
produce model colors.  For each object, the best-matching template
and, hence, redshift and spectral type, are chosen by $\chi^2$
minimization of the observed fluxes with respect to the model colors.

\subsection{Toronto 7-color}

The Toronto 7-color method is identical to the Toronto 4-color method
except that (a)~3 IR filters ($I$, $J$ and $K$; Dickinson et al.\
1997) are added and (b)~for the purposes of measuring galaxy colors,
the seeing in the 4 HST images is degraded to make them compatible
with the ground-based IR data.  The inclusion of the IR data decreases
the confusion between the various spectral breaks --- in particular
the Balmer and the 912\,\AA\ break --- and thereby increases the
accuracy of photometric redshifts.

\subsection{Johns Hopkins}

Because the uncertainties in using low-redshift spectral energy
distributions are large, particularly in the ultraviolet, the Johns
Hopkins method adopts a more empirical approach to the problem of
photometric redshift prediction.  Using recent spectroscopic redshift
survey data as a training set, an empirical relation between the
photometric properties of galaxies (their fluxes and colors) and their
redshifts is derived. For the HDF a correlation between the optical
and followup near-infrared $J$ band photometry (Dickinson et al 1997)
and redshift is derived.  A third-order polynomial in the $F300W$,
$F450W$, $F606W$, $F814W$ and $J$ passbands is fit to the 73 galaxies
with previously published spectroscopic data (Cohen et al 1996,
Steidel et al 1996, Lowenthal et al 1996).  This relation is applied
to those galaxies without spectroscopic redshifts.  The inclusion of
the near-infrared data significantly improves the precision of the
photometric redshift relation, primarily for $z>1$, because the
$\sim$4000\,\AA\ break moves out of the optical spectral region and
into the near-infrared at these redshifts. For star--forming galaxies,
the ultraviolet continuum from Lyman~$\alpha$ (1215\,\AA) to
$\sim$3000\,\AA\ is relatively devoid of strong features, and
consequently there is little information in the optical photometry
from which to estimate redshifts between $1 < z < 2$. Incorporating
the $J$ band alone extends the redshift interval over which reliable
photometric redshifts can be obtained to $z\sim 2$. Full details
discussing the derivation and application of this photometric redshift
technique can be found in Connolly et al (1995), SubbaRao et al
(1996), Brunner et al (1997) and Connolly et al (1997).

\subsection{Stony Brook}

The Stony Brook technique is essentially that of Lanzetta et al.\
(1996), but takes advantage of ground-based near-infrared images of
the HDF (Dickinson et al.\ 1997). A more detailed description of this
work is presented elsewhere (Fern\'andez-Soto et al.\ 1997).  Flux
measurements of every object in the HDF optical images are performed
using identical apertures on all four filters. For every object, its
aperture is given by a mask determined by the set of pixels that the
SExtractor program (Bertin \& Arnouts 1996) assigns to it in the F814W
image.  By convolving the F814W image of each source within its mask
aperture with the appropriate near-infrared ($J$, $H$ or $K$ band) PSF
and normalizing, a model of the infrared images can be created.  This
procedure makes the assumption that morphology is not a strong
function of wavelength.  The model has as parameters the fluxes of
every object in the $J$, $H$ and $K$ bands.  This model is fitted to
the data to provide matched near-infrared fluxes for every object
together with their errors.  These fluxes are compared to the expected
fluxes from the model SEDs (Lanzetta et al.\ 1996; the SEDs are
essentially those of Coleman, Wu \& Weedman 1980), extended to the
near-infrared according to the models of Bruzual and Charlot
(1993). Intergalactic HI absorption was taken into account as
described in Lanzetta et al.\ (1996).  No galaxy-star separation was
performed, but obviously stellar sources were removed after visual
inspection; this can lead to some misassignations of non-zero
redshifts to stars.

\section{Results}

Different groups make predictions for different total numbers of
sources because their catalogs are selected in different bands with
different algorithms over different areas.

As Table~\ref{tab:results} shows, all photometric-redshift techniques
surveyed here do well, with many sources in agreement with
spectroscopy at the $|\Delta z|<0.1$ level.  When only multi-feature
spectroscopic redshifts are considered, most techniques show better
than 68~percent agreement at the $|\Delta z|<0.1$ level and all show
90--100~percent agreement with spectroscopy at the $|\Delta z|<0.3$
level (when stars are removed from the Stony Brook catalog).
Unfortunately, the spectroscopic sample does not include sources past
$z=1.4$, so only lower redshifts are tested.  The results of this
study should be considered encouraging for those interested in using
photometric redshifts to define spectroscopic samples and estimate
luminosity functions at $0.4<z<1$, with the caveats that the
photometric data here are of higher-than-average quality and that it
will be necessary to perform studies with a larger number of redshifts
to make a rigorous test and to characterize systematics.  The results
are shown graphically in a redshift--redshift diagram in
Figure~\ref{fig:zz}, in a redshift-error--redshift diagram in
Figure~\ref{fig:dzz}, and in redshift-error histograms in
Figure~\ref{fig:dzhist}.  No strong systematic discrepancies between
photometric and spectroscopic redshifts are evident, although given
the small number of sources only systematic effects larger than about
0.2 in redshift over the tested redshift range could be detected.  For
all groups the 95-percent error region in $|\Delta z|$ is more than
twice the size of the 68-percent; this provides some evidence that the
discrepancies between photometric and spectroscopic redshifts are not
gaussian-distributed.

\acknowledgements

The HDF team, led by Bob Williams, are thanked for planning, taking,
reducing and making public the beautiful HST images of the HDF, as are
Len Cowie and the Hawaii group for confirming some originally
uncertain redshifts.  DWH thanks the Canadian Institute for
Theoretical Astrophysics for hospitality.  Support from the NSF and
NASA is gratefully acknowledged, including NSF grant AST-9529170, and
Hubble Fellowship grant HF-01093.01-97A from STScI (operated by AURA
under NASA contract NAS~5-26555).


\clearpage
\plotone{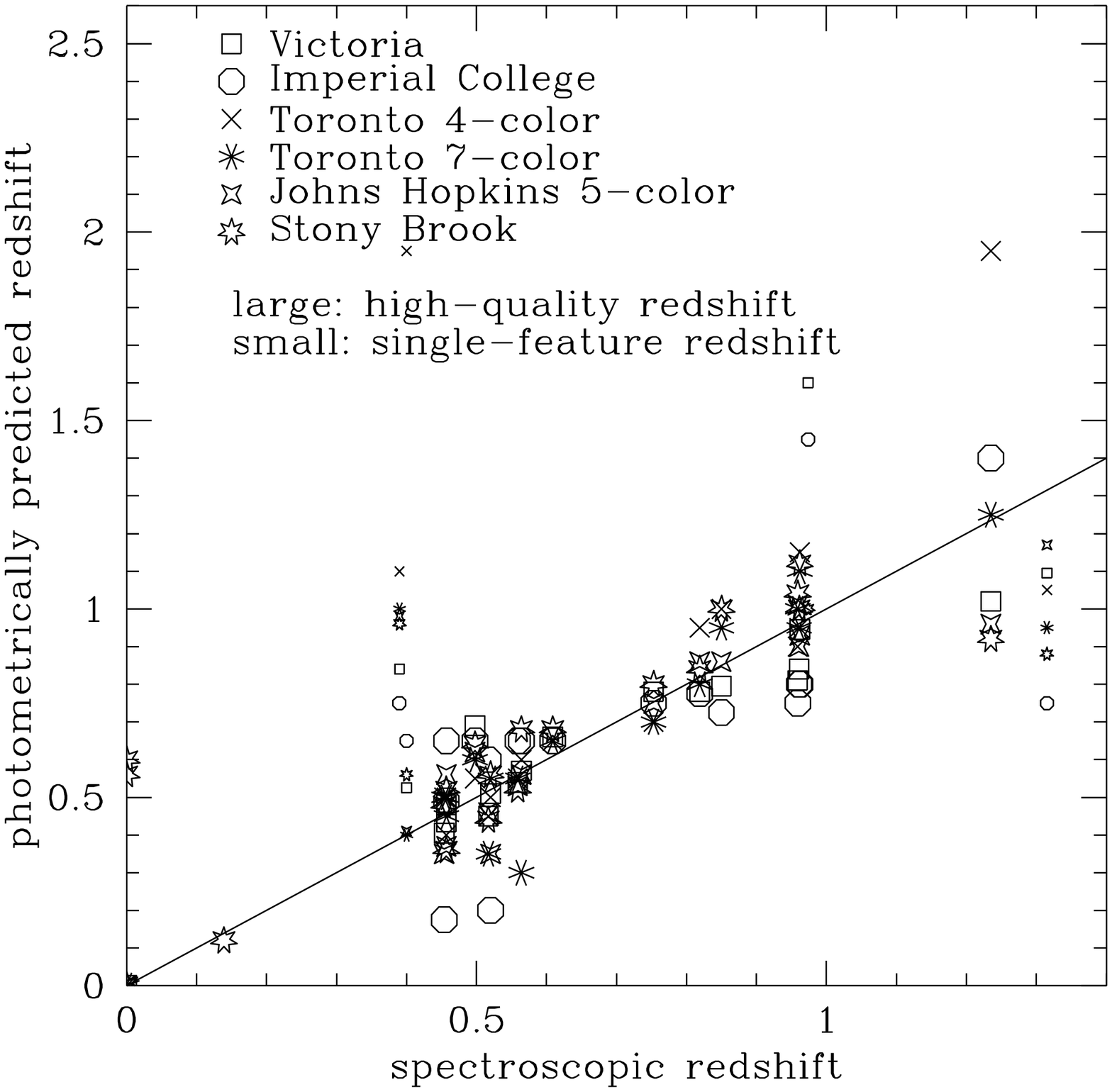}
\figcaption[Hogg.fig1.ps]{
Photometric versus spectroscopic redshifts for the sample.  Smaller
symbols are used for single-feature spectroscopic redshifts (see text).
\label{fig:zz}}

\clearpage
\plotone{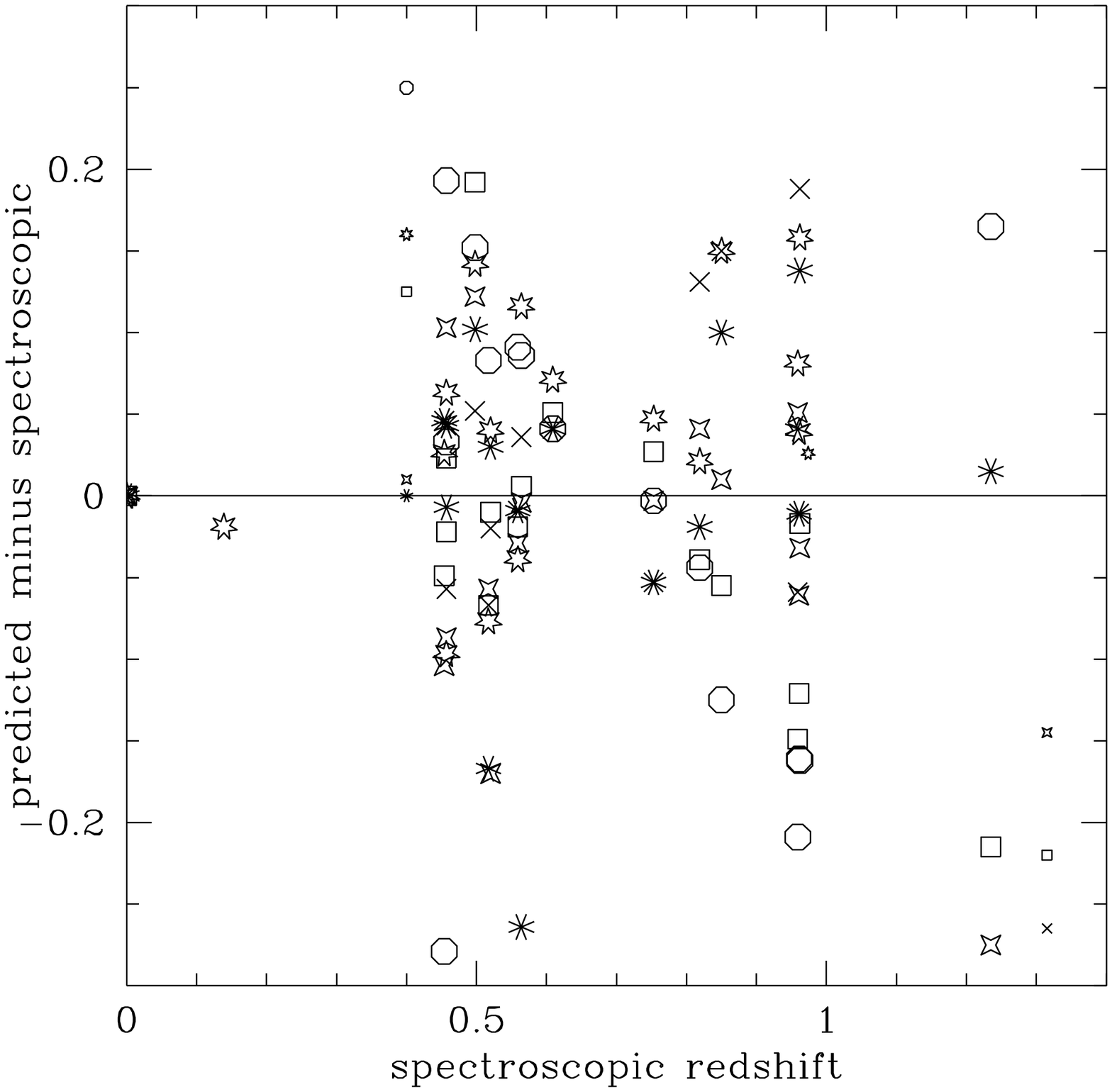}
\figcaption[Hogg.fig2.ps]{
Redshift differences, predicted minus spectroscopic, expanded to show
only those predictions with differences $|\Delta z|<0.3$.  Symbols as
in Figure~\ref{fig:zz}.
\label{fig:dzz}}

\clearpage
\plotone{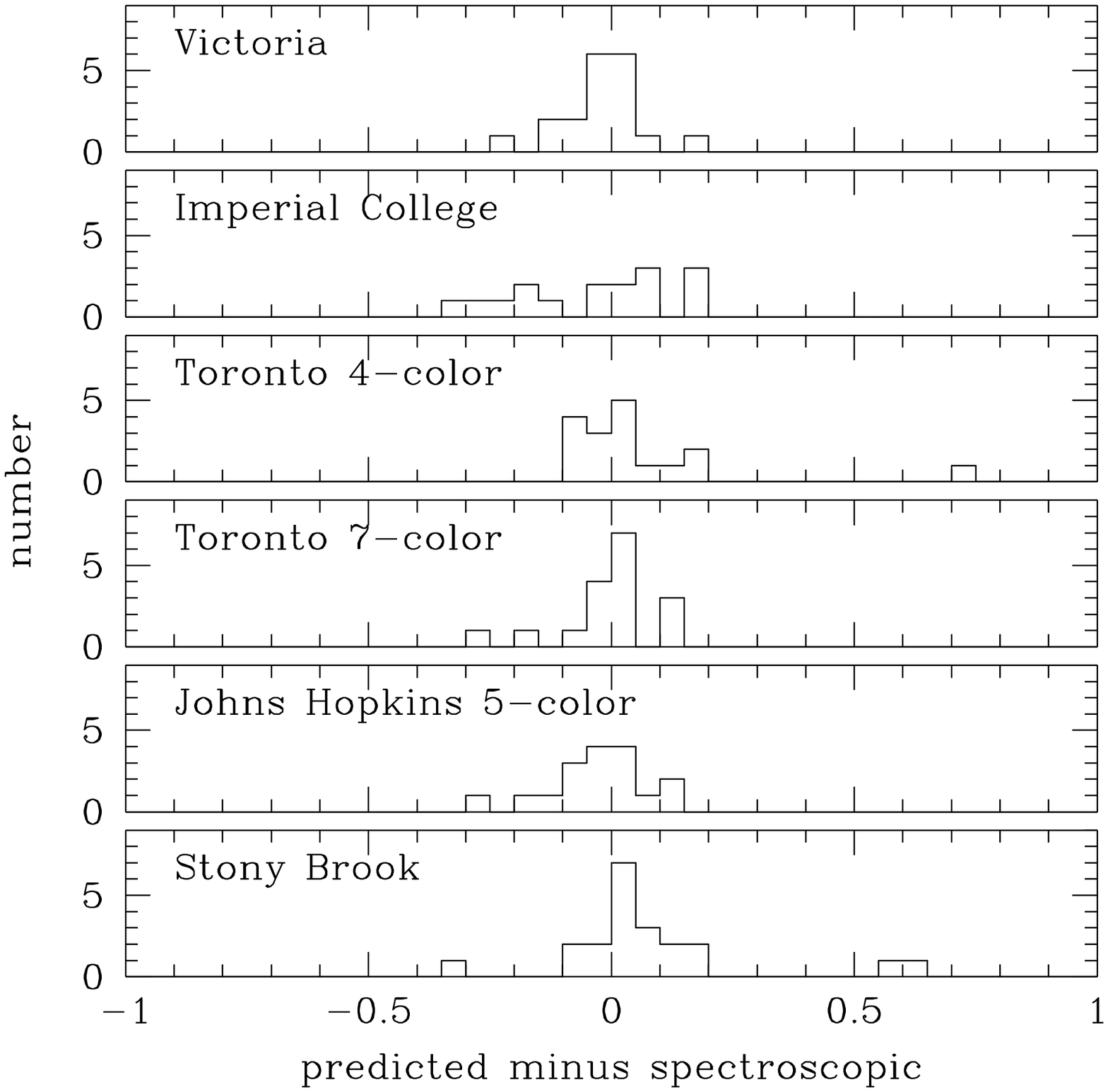}
\figcaption[Hogg.fig3.ps]{
Histogram of redshift differences for each group, excluding those sources
with single-feature spectroscopic redshifts.
\label{fig:dzhist}}

\clearpage
\begin{deluxetable}{lccccl}
\singlespace
\tablecaption{Sources}
\tablehead{
   \colhead{name\tablenotemark{a}}
 & \colhead{chip}
 & \colhead{$x,y$\tablenotemark{b}}
 & \colhead{$F606W$\tablenotemark{c}}
 & \colhead{$F606W-F814W$\tablenotemark{c}}
 & \colhead{$z$\tablenotemark{d}}
\\
 &
 & \colhead{(pix)}
 & \colhead{(mag)}
 & \colhead{(mag)}
}
\startdata
H36431\_1148 & 4 & 1554,1582 & 24.1 & 1.7 & ---\\
H36448\_1200 & 4 & 1385,1199 & 23.8 & 0.8 & 0.457\\
H36461\_1246 & 4 &  404,535  & 23.9 & 1.9 & ---\\
H36465\_1203 & 4 & 1437,890  & 24.7 & 0.8 & 0.454\\
H36469\_1422 & 2 & 2030,792  & 20.8 & 1.3 & 0.000\\
H36471\_1414 & 2 & 1831,753  & 24.5 & 1.4 & 0.609\\
H36477\_1232 & 4 &  831,400  & 24.2 & 1.2 & 0.959\\
H36483\_1426 & 2 & 2032,1060 & 19.8 & 0.6 & 0.139\\
H36483\_1214 & 4 & 1307,514  & 24.3 & 1.3 & 0.962\\
H36483\_1249 & 4 &  478,148  & 22.8 & 2.4 & 0.000\\
H36487\_1318 & 2 &  434,446  & 24.0 & 1.1 & 0.753\\
H36492\_1148 & 4 & 1961,599  & 24.2 & 1.6 & 0.961\\
H36503\_1418 & 2 & 1709,1307 & 24.2 & 1.2 & 0.819\\
H36508\_1250 & 3 &  521,634  & 24.3 & 0.8 & 0.40:\\
H36519\_1332 & 2 &  527,1096 & 24.0 & 1.3 & 0.39:\\
H36520\_1209 & 3 &  294,1667 & 23.6 & 0.8 & 0.457\\
H36520\_1400 & 2 & 1184,1394 & 23.8 & 1.0 & 0.559\\
H36528\_1405 & 2 & 1220,1574 & 24.2 & 1.2 & 0.498\\
H36536\_1417 & 2 & 1451,1826 & 24.1 & 0.9 & 0.517\\
H36540\_1245 & 3 &  980,981  & 24.1 & 1.9 & 0.000\\
H36541\_1354 & 2 &  886,1662 & 23.5 & 1.3 & 0.850\\
H36554\_1311 & 3 & 1463,489  & 24.1 & 2.1 & 1.315:\\
H36555\_1402 & 2 &  979,1971 & 24.2 & 1.3 & 0.564\\
H36567\_1252 & 3 & 1481,1009 & 25.1 & 1.7 & 1.235\\
H36569\_1258 & 3 & 1570,892  & 24.3 & 1.1 & 0.520\\
H36593\_1255 & 3 & 1928,1113 & 24.3 & 2.3 & 0.000\\
H37016\_1225 & 3 & 2009,1967 & 24.5 & 1.1 & 0.974:\\
\enddata
\tablenotetext{a}{Sources are named by the convention that, for
example the source at $12^h\,37^m\,01.6^s~+62^{\circ}\,12'\,25''$
(J2000) is called ``H37016\_1225''.}
\tablenotetext{b}{$x,y$ positions are given in pixels in the
appropriate chip in the ``Version 2'' reduction of the HST HDF data.}
\tablenotetext{c}{All magnitudes are measured through 1.5~arcsec
diameter apertures.}
\tablenotetext{d}{Redshifts $z$ are from Cohen et al (1997).  Sources
without a successfully measured spectroscopic redshift are indicated
with dashes.  Sources with $z=0.000$ are stars.  Redshifts based on a
single feature are marked with a colon.}
\label{tab:sources}
\end{deluxetable}

\clearpage
\begin{deluxetable}{llllllll}
\singlespace
\tablewidth{0pt}
\tablecaption{Redshift predictions}
\tablehead{
   \colhead{name}
 & \colhead{spec.}
 & \colhead{Victoria}
 & \colhead{Imperial}
 & \colhead{Toronto-4}
 & \colhead{Toronto-7}
 & \colhead{Hopkins}
 & \colhead{Stony Brook}
}
\startdata
H36431\_1148 & ---    & 0.900 & 0.600 & 1.05  & 1.10  & 1.22  & 1.04  \\
H36448\_1200 & 0.457  & 0.480 & 0.490 & 0.40  & 0.50  & 0.56  & 0.36  \\
H36461\_1246 & ---    & 0.765 & 0.675 & 0.90  & 0.65  & 0.75  & 0.60  \\
H36465\_1203 & 0.454  & 0.405 & 0.175 & 0.50  & 0.50  & 0.35  & 0.48  \\
H36469\_1422 & 0.000  & 0.000 & ---   & ---   & ---   & ---   & 0.000 \\
H36471\_1414 & 0.609  & 0.660 & 0.650 & 0.65  & 0.65  & ---   & 0.68  \\
H36477\_1232 & 0.959  & 0.810 & 0.750 & 0.90  & 1.00  & 1.01  & 1.04  \\
H36483\_1214 & 0.962  & 0.945 & 0.800 & 1.15  & 1.10  & 0.93  & 1.12  \\
H36483\_1249 & 0.000  & ---   & ---   & ---   & ---   & ---   & 0.000 \\
H36483\_1426 & 0.139  & ---   & ---   & ---   & ---   & ---   & 0.12  \\
H36487\_1318 & 0.753  & 0.780 & 0.750 & 0.70  & 0.70  & 0.75  & 0.80  \\
H36492\_1148 & 0.961  & 0.840 & 0.800 & 0.95  & 0.95  & 0.90  & 1.00  \\
H36503\_1418 & 0.819  & 0.780 & 0.775 & 0.95  & 0.80  & 0.86  & 0.84  \\
H36508\_1250 & 0.40:  & 0.525 & 0.650 & 1.95  & 0.40  & 0.41  & 0.56  \\
H36519\_1332 & 0.39:  & 0.840 & 0.750 & 1.10  & 1.00  & 0.98  & 0.96  \\
H36520\_1209 & 0.457  & 0.435 & 0.650 & 0.50  & 0.45  & 0.37  & 0.52  \\
H36520\_1400 & 0.559  & 0.540 & 0.650 & 0.55  & 0.55  & 0.53  & 0.52  \\
H36528\_1405 & 0.498  & 0.690 & 0.650 & 0.55  & 0.60  & 0.62  & 0.64  \\
H36536\_1417 & 0.517  & 0.450 & 0.600 & 0.45  & 0.35  & 0.46  & 0.44  \\
H36540\_1245 & 0.000  & 0.000 & ---   & 0.000 & 0.000 & 0.000 & 0.60  \\
H36541\_1354 & 0.850  & 0.795 & 0.725 & 1.00  & 0.95  & 0.86  & 1.00  \\
H36554\_1311 & 1.315: & 1.095 & 0.750 & 1.05  & 0.95  & 1.17  & 0.88  \\
H36555\_1402 & 0.564  & 0.570 & 0.650 & 0.60  & 0.30  & 0.56  & 0.68  \\
H36567\_1252 & 1.235  & 1.020 & 1.400 & 1.95  & 1.25  & 0.96  & 0.92  \\
H36569\_1258 & 0.520  & 0.510 & 0.200 & 0.50  & 0.55  & 0.35  & 0.56  \\
H36593\_1255 & 0.000  & 0.000 & ---   & ---   & ---   & 0.000 & 0.56  \\
H37016\_1225 & 0.974: & 1.600 & 1.450 & ---   & ---   & ---   & 1.00  \\
\enddata
\label{tab:predictions}
\end{deluxetable}

\clearpage
\begin{deluxetable}{lcccc}
\singlespace
\tablewidth{0pt}
\tablecaption{Results}
\tablehead{
 & \multicolumn{2}{c}{$|\Delta z|\leq 0.1$}
 & \multicolumn{2}{c}{$|\Delta z|\leq 0.3$}
\\
   \colhead{institution (submitter)}
 & \colhead{total}
 & \colhead{(good)\tablenotemark{a}}
 & \colhead{total}
 & \colhead{(good)\tablenotemark{a}}
}
\startdata
University of Victoria (Gwyn)                            & 15/23 & (15/19) & 21/23 & (19/19) \\
Imperial College (Mobasher)                              &  7/20 &  (7/16) & 16/20 & (15/16) \\
University of Toronto 4-color (Sawicki)                  & 13/20 & (13/17) & 17/20 & (16/17) \\
University of Toronto 7-color (Sawicki)\tablenotemark{b} & 14/20 & (13/17) & 18/20 & (17/17) \\
Johns Hopkins University (Connolly)\tablenotemark{b}     & 13/20 & (12/17) & 19/20 & (17/17) \\
SUNY Stony Brook (Fern\'andez-Soto)\tablenotemark{b,c}   & 15/25 & (14/21) & 20/25 & (18/21) \\
\enddata
\tablenotetext{a}{The ``good'' columns do not include single-feature
spectroscopic redshifts (see text).}
\tablenotetext{b}{Makes use of additional ground-based infrared data.}
\tablenotetext{c}{The Stony Brook results include 2 sources which are
stars but predicted to have redshifts $z\sim 0.6$.  Removing these
sources, the Stony brook numbers become 15/23 (14/19) 20/23 and
(18/19) respectively.}
\label{tab:results}
\end{deluxetable}


\begin{references}
\reference{}
Baum, W. A., 1962, {\it IAU Symp.\ 15:\ Problems of extra-galactic
research,} ed.\ McVittie, G. C., 390
\reference{}
Bertin, E. \& Arnouts, S., 1996, A\&A, 117, 393
\reference{}
Brunner, R.~J., Connolly, A.~J., Szalay, A.~S. \& Bershady, M.~A.,
1997, ApJ in press
\reference{}
Bruzual A., G. \& Charlot, S., 1993, ApJ, 405, 538
\reference{}
Bruzual A., G. \& Charlot, S. 1997, in preparation
\reference{}
Cohen, J. G., Cowie, L. L., Hogg, D. W., Songaila, A., Blandford, R.,
Hu, E. M. \& Shopbell, P., 1996, ApJ, 471, L5
\reference{}
Cohen, J. G. et al., 1997, in preparation
\reference{}
Coleman, G. D., Wu, C.-C., \& Weedman, D. W., 1980, ApJS, 43, 393
\reference{}
Connolly, A. J., Csabai, I., Szalay, A.~S., Koo, D.~C., Kron, R.~C. \&
Munn, J.~A., 1995, AJ, 110, 2655
\reference{}
Connolly, A. J., Szalay, A. S., Dickinson, M., SubbaRao, M. U. \&
Brunner, R. J., 1997, ApJ, 486, 11
\reference{}
Dickinson, M., et al., 1997, in preparation
\reference{}
Ferguson, H.~C. \& McGaugh, S.~S., 1994, ApJ, 440, 470
\reference{}
Fern\'andez-Soto, A., Lanzetta, K.M. \& Yahil, A., 1997, in preparation
\reference{}
Gwyn, S.~D.~J. \& Hartwick, F.~D.~A., 1996, ApJ, 468, L77
\reference{}
Gwyn, S.~D.~J., 1997, PASP, submitted
\reference{}
Hogg, D. W., Cohen, J. G. \& Blandford, R., 1997, poster at the May
1997 meeting at STScI on {\it The Hubble Deep Field}
\reference{}
Koo, D. C., 1985, AJ, 90, 418
\reference{}
Lanzetta, K.M., Yahil, A. \& Fern\'andez-Soto, A., 1996, Nature, 381, 759
\reference{}
Loh, E. D. \& Spillar, E. J., 1986, ApJ, 303, 154
\reference{}
Lowenthal, J.~D., Koo, D.~C., Guzman, R. Gallego, J., Phillips, A.~C.,
Vogt, N.~P., Faber, S.~M., Illingworth, G.~D. \& Gronwall, C., 1997,
ApJ, 481, 673
\reference{}
Madau, P., 1995, ApJ, 441, 18
\reference{}
Mazzei, P., De Zotti, G., Xu, C., 1994, ApJ 422, 81
\reference{}
Mazzei, P., Curir, A., Bonoli, C., 1995, AJ 110, 559
\reference{}
Mobasher, B., Rowan-Robinson, M., Georgakakis, A. \& Eaton, N., 1996,
MNRAS, 282, L7
\reference{}
Mobasher, B., Mazzei, P. \& Rowan Robinson, M., 1997, MNRAS, in
preparation
\reference{}
Sawicki, M. J., Lin, H., \& Yee, H. K. C., 1997, AJ, 113, 1
\reference{}
Steidel, C.~C., Giavalisco, M., Dickinson, M. \& Adelberger, K.~L.,
1996, ApJ, 462, 17
\reference{}
SubbaRao, M.~U., Connolly A.~J., Szalay, A.~S. \& Koo, D.~C., 1996,
AJ, 112, 929
\reference{}
Williams R. E. et al, 1996, AJ, 112, 1335
\end{references}
\end{document}